\documentclass[preprint,11pt,nofootinbib]{revtex4-1}

\usepackage{graphicx}
\usepackage{dcolumn} 
\usepackage{bm}
\usepackage{bbm}
\usepackage{amssymb}
\usepackage{amsmath}
\usepackage{epsfig}    
\usepackage{color}
\usepackage{slashed}
\usepackage{hyperref}

\begin{document}

\title{Probing WIMPs in space-based gravitational wave experiments}

\author{Bo-Qiang Lu}
\email{bqlu@zjhu.edu.cn}
\affiliation{School of Science, Huzhou University, Huzhou, Zhejiang 313000, China}

\author{Cheng-Wei Chiang}
\email{chengwei@phys.ntu.edu.tw}
\affiliation{Department of Physics, National Taiwan University, Taipei, Taiwan 10617, Republic of China}
\affiliation{Physics Division, National Center for Theoretical Sciences, Taipei, Taiwan 10617, Republic of China}

\author{Da Huang}
\email{dahuang@bao.ac.cn}
\affiliation{National Astronomical Observatories, Chinese Academy of Sciences, Beijing, 100012, China}
\affiliation{School of Fundamental Physics and Mathematical Sciences, Hanzhou Institute for Advanced Study, UCAS, Hanzhou 310024, China}
\affiliation{International Centre for Theoretical Physics Asia-Pacific, Beijing/Hanzhou, China}


\begin{abstract}
    Although searches for dark matter have lasted for decades, no convincing signal has been found without ambiguity in underground 
detections, cosmic ray observations, and collider experiments. We show by example that gravitational wave (GW) observations can be a 
supplement to dark matter detections if the production of dark matter follows a strong first-order cosmological phase transition. 
We explore this possibility in a complex singlet extension of the standard model with CP symmetry.  We demonstrate three benchmarks in which the GW signals from the first-order phase transition are loud enough for future space-based GW observations, for example, BBO, U-DECIGO, LISA, Taiji, and TianQin.  While satisfying the constraints from the XENON1T experiment and the Fermi-LAT gamma-ray observations, the dark matter candidate with its mass around $\sim 1$~TeV in these scenarios has a correct relic abundance obtained by the Planck observations of the cosmic microwave background radiation.
\end{abstract}



\maketitle

{\it Introduction--}
Although mounting evidence collected from cosmological~\cite{Planck2016} and astrophysical~\cite{Rubin1970APJ} observations suggests that 
the dominant component of matter in our universe is the dark matter (DM), its nature still remains a myth even after about eighty years since 
its first discovery.  Among various DM particle scenarios, one of the most promising DM candidates is the weakly interacting massive particles 
(WIMPs), which have their mass at the electroweak scale and their coupling strength close to that of the Standard Model (SM) electroweak 
sector~\cite{Lee1977,Hut1977}.  In this scenario, WIMPs become non-relativistic and freeze out from the thermal bath when the temperature 
of the Universe drops below the WIMP mass during the cosmological expansion.

In recent years, first-order cosmological phase transitions in the early Universe have attracted lots of attention because these violent 
phenomena could induce intense gravitational wave (GW) radiations~\cite{Weir2018PTRSLA,Cai2022} which may be detected by ongoing and upcoming 
GW experiments.  If the first-order phase transition occurs at the electroweak scale, the peak frequency of the associated GWs would be 
red-shifted to the mHz band, falling right within the detectable range of many space-based GW experiments, such as LISA~\cite{LISA2017}, 
Taiji~\cite{Hu2017NSR,Ruan2020NA}, TianQin~\cite{Luo2016CQG,Hu2018CAG}, BBO~\cite{Crowder2005PRD}, and DECIGO~\cite{Seto2001PRL}. 
Although there exist strong physical motivations for the electroweak phase transition 
(EWPT) to be first-order~\cite{Kuzmin1985PLB,Cohen1990PLB,Morrissey2012NJP}, lattice computations indicate that the electroweak (EW) gauge 
symmetry breaking with a 125-GeV Higgs boson proceeds merely via a crossover transition at a temperature of $T\sim 100$ GeV~\cite{Onofrio2014PRL}.

In this work, we are interested in a complex singlet scalar extension of the SM that can provide a WIMP DM candidate and generate a sufficiently 
strong first-order EWPT~\cite{Chiang2021JCAP,Alanne2020JHEP}. In particular, we focus on the so-called Type-II EWPT~\cite{Cline2013JCAP} in this singlet-extended model, 
in which the first-order EWPT takes place via two steps as $(0,0) \to (0,~w) \to (v,~0)$, where the first and second variables in the 
parenthesis represent the vacuum expectation values (VEVs) of the SM Higgs doublet and the new scalar singlet, respectively. Since the singlet 
VEV vanishes after the phase transition, one can make this extra scalar a DM candidate if we further impose an additional symmetry to ensure 
its stability.  Some popular attempts along this direction include the singlet scalar extensions with a 
$Z_2$~\cite{Cline2013JCAP,Guo2010JHEP,Musolf2019} or a $U(1)$~\cite{Hashino2018JHEP,Kannike2019PRD} symmetry.  
However, it has been shown that the type-II EWPT as well as the correct DM relic abundance in these models
favor a large value of the Higgs portal coupling $\lambda_m$, which however suffers from the strong constraints 
from the DM relic density and direct detections~\cite{Feng2015JHEP,Chiang2021JCAP}. Under the severe updated constraints from LUX and 
XENON1T, only a negligible fraction of ${\cal O}(\sim 10^{-4} - 10^{-5})$ of cosmological mass density can be composed of the singlet 
scalar DM~\cite{Chiang2021JCAP}.

We will focus on the GW phenomena induced by the first-order phase transition in the complex singlet extension of the SM imposed with CP 
symmetry $S\to S^*$ in the scalar potential, as proposed in our previous paper~\cite{Chiang2021JCAP}.  
We will demonstrate that in addition to providing a suitable WIMP candidate that renders a correct relic density while satisfying current 
DM detection bounds, the GW signals generated from the first-order EWPT in this model are sufficiently significant to be detectable 
by future space-based GW experiments.

{\it The model and phase transition -- }
We consider a complex singlet $S$ extension of the SM with the ``{\it CP} symmetry'' $S\to S^{*}$, under which the most general renormalizable scalar potential is given by
\begin{eqnarray}
    \label{eq:cppoten}
    V(H,S)=&-&\mu_{h}^2|H|^2+\lambda_h|H|^4-\mu_1^2(S^*S)-\frac{1}{2}\mu_2^2(S^2+S^{*2})+\lambda_1(S^*S)^2+\frac{1}{4}\lambda_2(S^2+S^{*2})^2\nonumber\\
    &+&\frac{1}{2}\lambda_3(S^*S)(S^2+S^{*2})+\kappa_1|H|^2(S^*S)+\frac{1}{2}\kappa_2|H|^2(S^2+S^{*2})+\frac{1}{\sqrt{2}}a_1^3(S+S^{*})\nonumber\\
    &+&\frac{1}{2\sqrt{2}}b_m|H|^2(S+S^{*})+\frac{\sqrt{2}}{3}c_1(S^*S)(S+S^*)+\frac{\sqrt{2}}{3}c_2(S^3+S^{*3})
    ~,
\end{eqnarray}
where all parameters are assumed to be real.  After a trivial rephasing, the Higgs doublet and singlet scalar fields can be expanded as
\begin{equation}
    H=\left(\begin{array}{c}
    G^{+} \\
    \frac{1}{\sqrt{2}}\left(h+i G^{0}\right)
    \end{array}\right)\,, \quad {\rm and}\quad S=\frac{1}{\sqrt{2}}(s+i \chi)
    \,.
\end{equation}
After the EW gauge symmetries are spontaneously broken, $G^{\pm}$ and $G^0$ become the longitudinal modes of the massive $W^\pm$ and $Z$ bosons, respectively, while $\chi$ is protected by the residual $CP$ symmetry so that it can be a DM candidate.  
In terms of the background fields, the effective scalar potential at finite temperature is given by
\begin{equation}\label{eq:potentot}
  V_{\mathrm{eff}}(h,s,\chi,T)=V_0(h,s,\chi)+V_{\mathrm{CW}}(h,s,\chi)+V_{T}(h, s,\chi,T)
  ~,
\end{equation}
where the tree-level potential at zero temperature is 
\begin{equation}\label{eq:poten2}
    \begin{aligned}
    V_{0}(h, s, \chi)=&-\frac{1}{2} \mu_{h}^{2} h^{2}-\frac{1}{2} \mu_{s}^{2} s^{2}-\frac{1}{2} \mu_{\chi}^{2} \chi^{2}+\frac{1}{4} 
    \lambda_{h} h^{4}+\frac{1}{4} \lambda_{s} s^{4}+\frac{1}{4} \lambda_{\chi} \chi^{4}+\frac{1}{2} \lambda_{a} s^{2} \chi^{2} \\
    &+\frac{1}{4} \kappa_{s} h^{2} s^{2}+\frac{1}{4} \kappa_{\chi} h^{2} \chi^{2}+a_{1}^{3} s+\frac{1}{4} b_{m} h^{2} s+\frac{1}{3} 
    c_{s} s^{3}+\frac{1}{3} c_{\chi} s \chi^{2}
    ~.
    \end{aligned}
\end{equation}
The parameters appearing in $V_0$ are related to those used in Eq.~\eqref{eq:cppoten}. Please refer to Ref.~\cite{Chiang2021JCAP} for more details.
The one-loop corrections are given by the Coleman-Weinberg potential~\cite{Coleman1973PRD}
\begin{equation}
V_{\mathrm{CW}}(h,s,\chi)=\frac{1}{64 \pi^{2}} \sum_{i} N_{i} M_{i}^{4}(h, s,\chi)\left[\log \frac{M_{i}^{2}(h,s,\chi)}{\mu^{2}}-C_{i}\right]
~,
\end{equation}
where the subscript $i$ runs over $s$, $\chi$, and the SM particles, $M_{i}$ is the field-dependent bosonic mass, $N_i$ is the number of degrees of freedom of the particles, $\mu$ is renormalization scale, and the constant $C_{i}=1 / 2$ for gauge boson transverse modes and $3 / 2$ for all the other particles. Please refer to Refs.~\cite{Chiang2020JHEP,Coleman1973PRD} for more details on the Coleman-Weinberg potential.
The finite-temperature contributions to the effective potential at one-loop level are given by
\begin{equation}
V_{\mathrm{T}}(h, s,\chi, T)=\frac{T^{4}}{2 \pi^{2}} \sum_{i} N_{i} J_{B, F}\left(M_{i}^{2}(h, s,\chi, T) / T^{2}\right)
~,
\end{equation}
where $J_{B, F}\left(z^{2}\right)=\int_{0}^{\infty} d x x^{2} \ln \left(1 \mp e^{-\sqrt{x^{2}+z^{2}}}\right)$, with the $-$ sign for bosons 
and $+$ for fermions. In the high-temperature expansion, the leading-order one-loop finite-temperature corrections are given by
\begin{equation}\label{eq:VT}
   V_T(h,s,\chi,T)= \frac{1}{2}\left(g_{h} h^{2}+g_{s} s^{2}+g_{\chi} \chi^{2}+2 m_{3} s\right)T^{2}
   ~.
\end{equation}
The parameters in this equation can be found in Ref.~\cite{Chiang2021JCAP}. Note that the VEV of the field $\chi$ is assumed to be zero 
throughout the EWPT. The high temperature expansion at the leading order leads to a 
gauge-invariant potential~\cite{Patel2011JHEP,Chiang2018PRD}.

\begin{figure}
    \centering
    \includegraphics[width=55mm,angle=0]{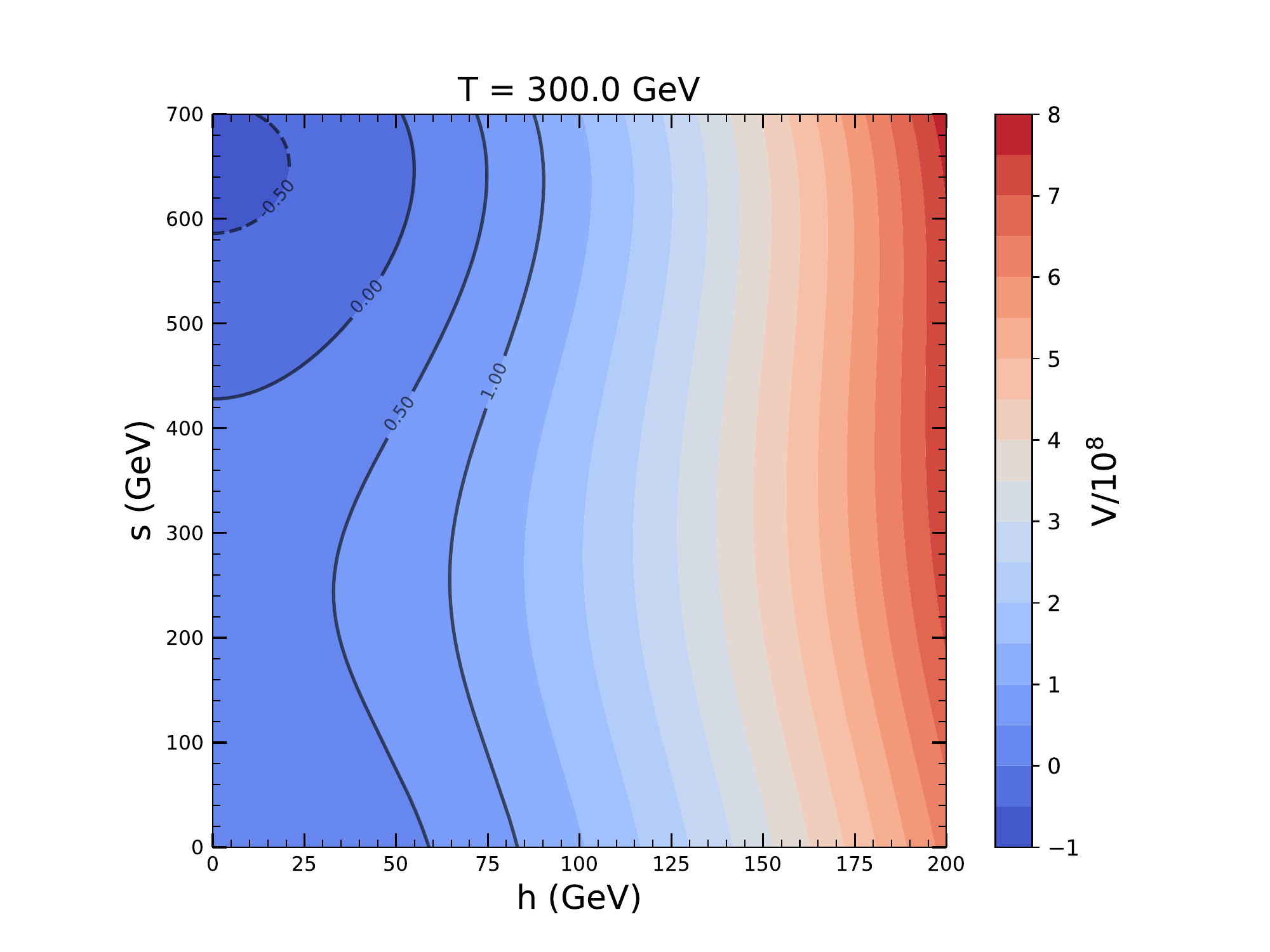}
    \hspace{-10mm}
    \includegraphics[width=55mm,angle=0]{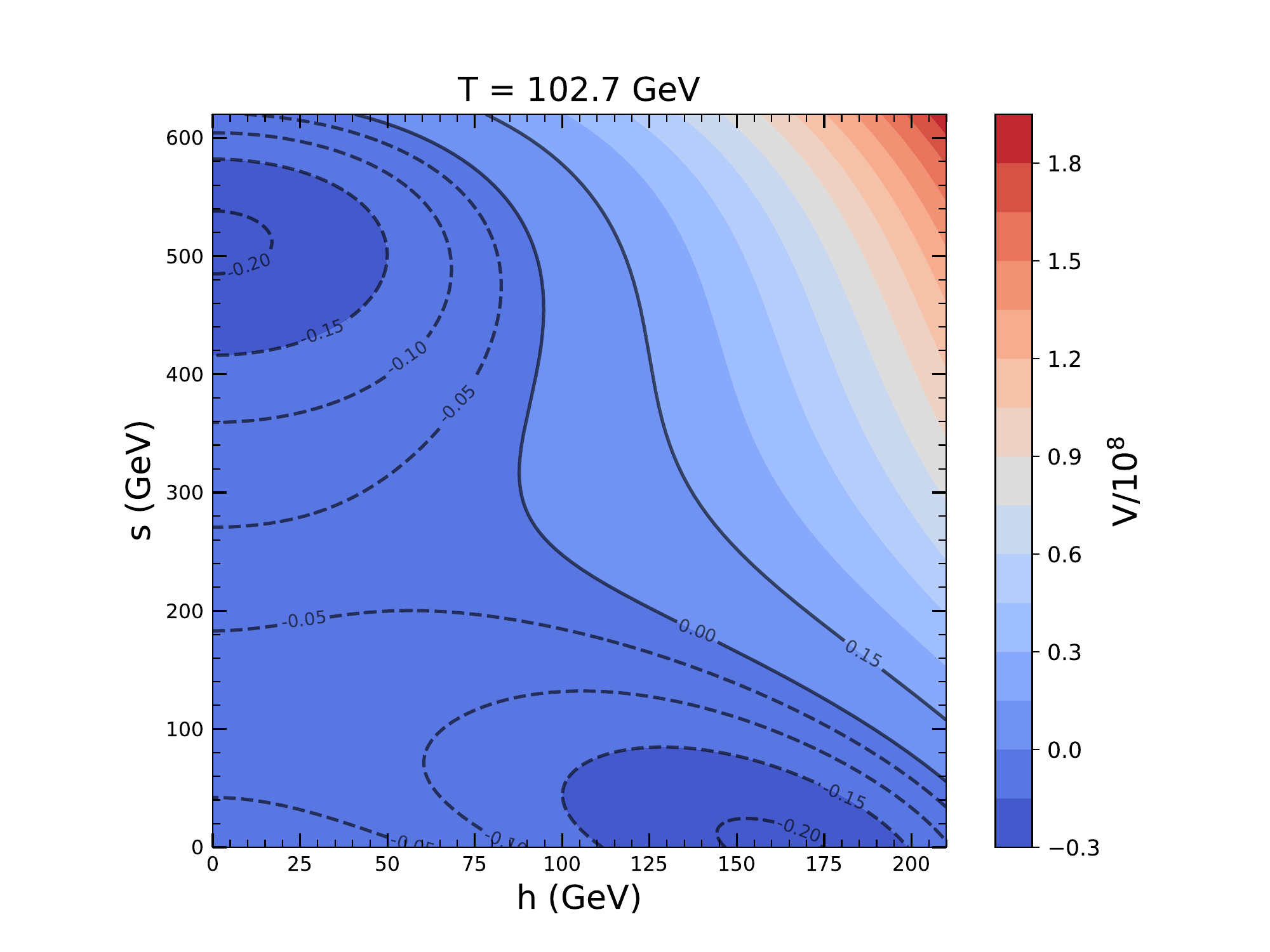}
    \hspace{-10mm}
    \includegraphics[width=55mm,angle=0]{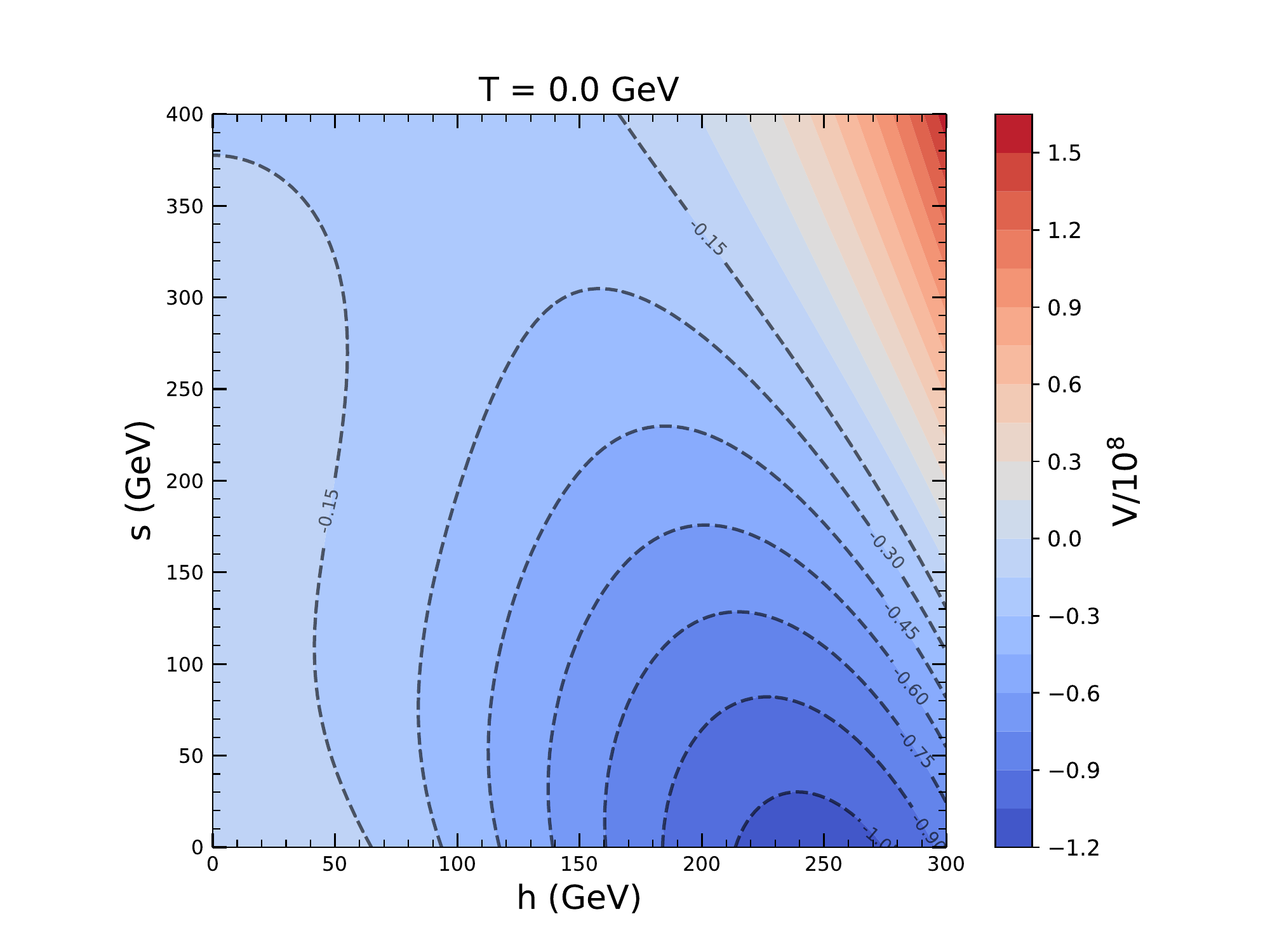}
    \caption{Contours of the effective potential in the $h-s$ plane, with temperature $T=300$~GeV (left), $T=T_c$ (middle), and $T=0$~GeV (right), respectively. 
    Here we take the parameters $\lambda_{\chi}=0.5$, $\kappa_{\chi}=0$, $m_S=48.9$~GeV,
    $\theta=-0.258$, $\lambda_s=1.54\times 10^{-2}$, $\lambda_a=5.25\times 10^{-3}$, $k_{s}=6.31\times 10^{-3}$, $c_s=-12.92$~GeV, and $c_{\chi}=32.36$~GeV.
    The other parameters can be derived from Eq.~(3.14) of Ref.~\cite{Chiang2021JCAP}.
    }
    \label{fig:poten}
\end{figure}

As an illustration, we depict the type-II phase transition of the potential in Fig.~\ref{fig:poten}. 
In this plot, we adopt the approximation of high-temperature expansion. 
In this case, the model parameters should satisfy the relations given in Eq.~(3.14) of Ref.~\cite{Chiang2021JCAP} to successfully trigger a type-II EWPT. 
As shown in this figure, the scalar potential along the $s$ direction first develops a global minimum at high temperatures.  As the temperature 
drops to the critical temperature $T_c$, another local minimum located at $(v_c,0)$, designating the EW symmetry-broken phase, appears and 
becomes degenerate with the EW symmetric one $(0,w_c)$.  The broken phase $(v_0\equiv 246~{\rm GeV},0)$ becomes the global minimum as the 
temperature approaches zero since the decrease of potential at the EW symmetric phase is much slower than that at the EW broken phase.

In the following, we make use of the scenarios given in Ref.~\cite{Chiang2020JHEP} to search for the parameter space of first-order EWPT with the full one-loop effective potential~\eqref{eq:potentot}. Note that 
the parameters should be chosen so as to guarantee several theoretical bounds, including the tree-level vacuum stability~\cite{Gonderinger2012PRD},
perturbativity, and perturbative unitarity~\cite{Kanemura2016NPB}.
The most relevant experimental constraints are from the Higgs signal strength observations.  We further require the $h-s$ mixing angle $|\theta|\lesssim 0.4$ to be in agreement with these constraints~\cite{Chiang2020JHEP}.

{\it Nucleation temperature -- }
The DM phenomenology of the model has been studied in detail in Ref.~\cite{Chiang2021JCAP}.  In this work, we focus on the stochastic gravitational waves emitted from the strong first-order EWPT of the model.  A first-order cosmological phase transition begins when the Universe cools down to the critical temperature $T_c$, at which point a potential barrier separates the two degenerate phases, the EW symmetry-broken and -unbroken vacua.  The stochastic bubble nucleation 
takes place when the metastable vacuum successfully tunnels into the stable one with the help of thermal fluctuations.  The nucleation rate per unit volume and per unit time is approximately given by~\cite{Apreda2002NPB}
\begin{equation}
    \Gamma(T)=T^{4}\left(\frac{S_{3}}{2 \pi T}\right)^{\frac{3}{2}} e^{-\frac{S_{3}}{T}}
    ~,
\end{equation}
where the three-dimensional on-shell Euclidean action for a spherical bubble configuration can be written as follows
\begin{equation}
    S_{3}(T)=4 \pi \int d r r^{2}\left[\frac{1}{2}\left(\frac{d \phi}{d r}\right)^{2}+V_{\mathrm{eff}}(\phi, T)\right]
    ~.
\end{equation}
In order to obtain the Euclidean action, we need to know the bubble profile which can be determined by numerically solving the following equation of motion
\begin{equation}
    \frac{{d}^{2} \phi}{{d} r^{2}}+\frac{2}{r} \frac{{d} \phi}{{d} r}=\frac{\partial V_{\mathrm{eff}}}{\partial \phi}, 
    \quad{\rm with}\quad \phi^{\prime}(0)=0 \quad {\rm and}\quad \phi(\infty)=0
    ~.
\end{equation}
We employ the {\tt CosmoTransitions 2.0.2} package~\cite{CosmoTransitions2012} to perform such numerical calculations of the bubble profile and Euclidean action.

The temperature $T_*$ at which the GWs are most violently generated from the first-order phase transition can be approximated to be the nucleation temperature $T_n$, $T_*\simeq T_n$.  Note that the bubble nucleation proceeds efficiently when the bubbles are not diluted by the expansion of the Universe.  Thus, we can determine the nucleation temperature $T_n$ at which the bubble nucleation rate catches up with the 
Hubble expansion rate, i.e., $\Gamma(T_n)\simeq H(T_n)^4$. Here in a radiation dominated Universe the Hubble parameter is given by $H=1.66 g_{*}^{1/2}T^{2}/M_{\mathrm{pl}}$, with $g_{*} \simeq 110$ and $M_{\mathrm{pl}}=1.22 \times 10^{19} ~\mathrm{GeV}$.  This condition is well approximated by $S_3(T_n)/T_n\simeq 140$ for the phase transition occuring at the EW-scale temperature of $T\sim \mathcal{O}(100)$~GeV.  The criterion of bubble nucleation determines whether the first-order cosmological phase transition has successfully taken place or not.

{\it Gravitational waves -- }
In order to determine the stochastic GW spectrum produced from a first-order phase transition, we need to calculate the 
following two characteristic parameters~\cite{Kamionkowski1994PRD} 
\begin{equation}
    \alpha=\frac{\epsilon\left(T_{*}\right)}{\rho_{\operatorname{rad}}\left(T_{*}\right)} ~\text { and }~
    \frac{\beta}{H_{*}}=\left.T_{*} \frac{d}{d T}\left(\frac{S_{3}(T)}{T}\right)\right|_{T=T_{*}}
    ~,
\end{equation}
where $\rho_{\mathrm{rad}}=\pi^{2} g_{*} T^{4} / 30$ is the radiation energy density in the plasma, and the latent heat associated with the phase transition is $\epsilon(T)=T \partial \Delta V_{b s}(T)/\partial T-\Delta V_{b s}(T)$, with $\Delta V_{b s}(T)$ the potential 
difference between the EW broken and symmetric phases.  The parameter $\alpha$ characterizes the strength of a phase transition. Usually, $\alpha \lesssim 1$ for a typical phase transition without a large amount of supercooling.  The parameter $\beta$ defines the time variation of the nucleation rate, and thus $\beta/H_*$ dictates the characteristic frequency of the GW spectrum.

It has been shown that bubble collisions, sound waves and turbulence produced after the bubble collisions can be the sources of GW radiations during the percolation of bubbles, leading to the total GW spectrum~\cite{Caprini2016JCAP} $h^{2}\Omega_{\mathrm{GW}}\simeq h^{2}\Omega_{\mathrm{col}}+h^{2}\Omega_{\mathrm{sw}}+h^{2}\Omega_{\mathrm{turb}}$.  However, recent studies indicate that bubble collisions provide a negligible contribution to the final GW signal
as very little energies are deposited in the bubble walls. In what follows, we will restrict ourselves to the case of non-runaway bubbles, where the GWs can be effectively produced by the sound waves and turbulence.

The GW spectra from sound waves and turbulence are summarized as follows~\cite{Caprini2016JCAP,Cai2017JCAP}:
\begin{equation}
    h^{2} \Omega_{\mathrm{sw}}(f)=2.65 \times 10^{-6}\left(\frac{H_{*}}{\beta}\right)\left(\frac{\kappa_{\mathrm{sw}} \alpha}
    {1+\alpha}\right)^{2}\left(\frac{100}{g_{*}}\right)^{\frac{1}{3}} v_{w}\left(\frac{f}{f_{\mathrm{sw}}}\right)^{3}
    \left(\frac{7}{4+3\left(f / f_{\mathrm{sw}}\right)^{2}}\right)^{7 / 2}
    ~,
\end{equation}
\begin{equation}
    h^{2} \Omega_{\mathrm{turb}}(f)=3.35 \times 10^{-4}\left(\frac{H_{*}}{\beta}\right)\left(\frac{\kappa_{\mathrm{turb}} 
    \alpha}{1+\alpha}\right)^{\frac{3}{2}}\left(\frac{100}{g_{*}}\right)^{1 / 3} v_{w} \frac{\left(\frac{f}{f_{\mathrm{turb}}}\right)^{3}}
    {\left(1+\frac{f}{f_{\mathrm{turb}}}\right)^{\frac{11}{3}}\left(1+\frac{8 \pi f}{H_{0}}\right)}
    ~,
\end{equation}
where $f_{\mathrm{sw}}=1.9 \times 10^{-2} \mathrm{mHz} \frac{1}{v_{w}}\left(\frac{\beta}{H_{*}}\right)\left(\frac{T_{*}}{100 \mathrm{GeV}}\right)\left(\frac{g_{*}}{100}\right)^{\frac{1}{6}}$ is the red-shifted peak frequency of the GW spectrum from sound waves and $f_{\text {turb }}=2.7 \times 10^{-2} \mathrm{mHz}\frac{1}{v_{w}}\left(\frac{\beta}{H_{*}}\right)\left(\frac{T_{*}}{100\mathrm{GeV}} \right)\left(\frac{g_{*}}{100}\right)^{\frac{1}{6}}$ is the peak frequency for turbulence, $H_{0}=16.5 \times 10^{-3} \mathrm{mHz}\left(\frac{T_{*}}{100 \mathrm{GeV}}\right)\left(\frac{g_{*}}{100}\right)^{\frac{1}{6}}$ is the red-shifted Hubble constant, $v_w$ denotes the bubble wall velocity and is assigned to be around unity.  The efficiency factors $\kappa_{\mathrm{sw}}$ and $\kappa_{\mathrm{turb}}$ indicate respectively the fractions of latent heat that are transformed into the bulk motion of the plasma and the turbulence, in both of which the energy density finally goes into the emission of GWs.  For non-runaway bubbles with $v_{w} \sim 1$, the efficiency factors satisfy $\kappa_{\mathrm{sw}} \simeq \alpha/(0.73+0.083 \sqrt{\alpha}+\alpha)$ and $\kappa_{\text {turb }}=\epsilon \kappa_{\text {sw }}$, where we take $\epsilon=0.1$ in this work.  Note that the amplitude of the GW spectrum visible today is further suppressed by a factor of 
$\Upsilon=1-1/\sqrt{1+2 \tau_{\mathrm{sw}} H_{s}}$~\cite{Guo2021JCAP}.  We recommend Refs.~\cite{Guo2021JCAP,Hindmarsh2021SPLN,Ellis2021JCAP} 
for more details about recent developments toward understanding the GW production from first-order cosmological phase transitions.

{\it Signal-to-noise ratio -- }
With the frequentist approach, the detectability of the stochastic GW signals is measured by the corresponding signal-to-noise ratio (SNR)~\cite{Caprini2016JCAP}
\begin{equation}
    \rho=\sqrt{\mathcal{N} \mathcal{T}_{\mathrm{obs}} \int_{f_{\min }}^{f_{\max }} d f\left[\frac{h^{2} 
    \Omega_{\mathrm{GW}}(f)}{h^{2} \Omega_{\exp }(f)}\right]^{2}}
    ~,
\end{equation}
where $\mathcal{N}$ is the number of independent observatories of an experiment. For example, $\mathcal{N}=1$ for the auto-correlated experiments like LISA~\cite{LISA2017,Robson2019CQG} and B-DECIGO~\cite{Seto2001PRL,Isoyama2018PTEP}, while $\mathcal{N}=2$ for the cross-correlated experiments such as DECIGO~\cite{Sato2017JPCS} and BBO~\cite{Crowder2005PRD}. $\mathcal{T}_{\text {obs }}$ is the duration of the mission in units of year.  Here we assume a mission duration of $\mathcal{T}_{\text {obs }}=4$ years for all of the experiments.  $h^2\Omega_{\rm exp}$ denotes the sensitivity of a GW experiment, which is summarized in appendix~E of Ref.~\cite{Chiang2020JHEP}, where we also give the associated experimental frequency ranges ($f_{\rm min}-f_{\rm max}$)~Hz in Table~2.  We adopt the commonly used SNR threshold $\rho_{\rm thr}=10$, above which the GW signal can be regarded as detectable by one experiment.

\begin{figure}
    \centering
    \includegraphics[width=75mm,angle=0]{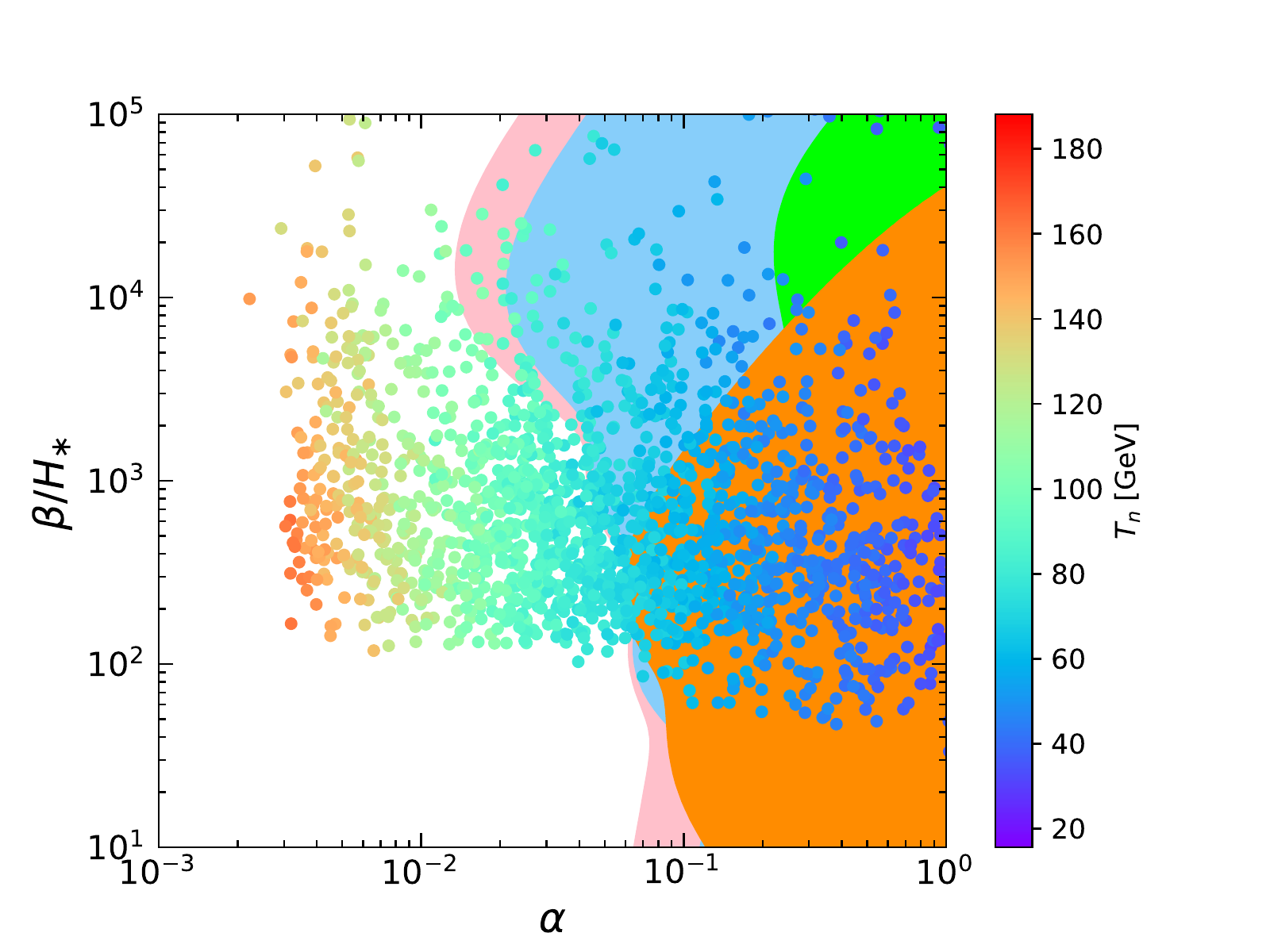}
    \hspace{-5mm}
    \includegraphics[width=75mm,angle=0]{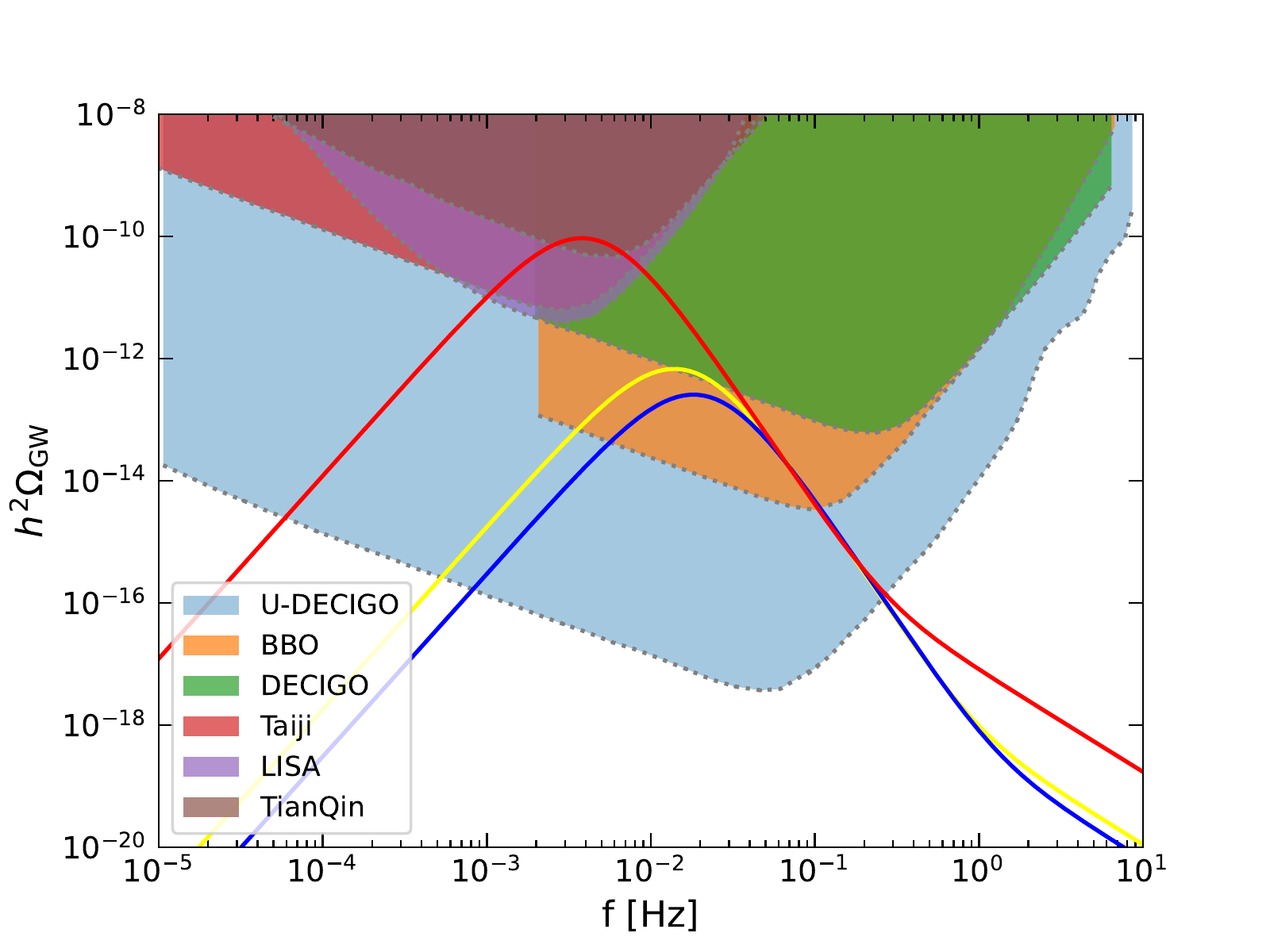}
    \caption{Left: the colored regions represent the parameter space with a SNR $\rho>\rho_{\rm thr}$ for LISA (orange), B-DECIGO (light-green), DECIGO (light-blue), and BBO (pink) experiments. The scatter points denote the scenarios that can both generate a strong first-order EWPT and have a correct DM relic density $h^2\Omega_{\rm DM}\simeq 0.12$.  The colorbar for the scatters denotes the nucleation temperature $T_n$, in units of GeV. 
    Right: GW spectra from the phase transition (curves) and various experimental sensitivities (colored patches) as a function of frequency.  The yellow, blue, and red curves represent the results in scenarios A, B, and C as summarized in Table~\ref{tab:models}.  Furthermore, these scenarios give a correct DM relic density 0.12.  
    Scenario A with a spin-independent DM-nucleon elastic scattering cross section $\sigma_{\rm SI}=1.27\times 10^{-46}$~$\rm cm^2$ 
    and scenario B with $\sigma_{\rm SI}=1.23\times 10^{-47}$~$\rm cm^2$ survive the XENON1T~\cite{XENON1T2018PRL} and PandaX~\cite{PandaX-II2017PRL} 
    constraints and are testable in the future XENONnT experiment, as well as the BBO and U-DECIGO GWs detections. 
    Scenario C with $\sigma_{\rm SI}=5.70\times 10^{-48}$~$\rm cm^2$ is beyond the future XENONnT~\cite{XENONnT2016JCAP} experimental sensitivity, 
    however its GW signal is detectable for the LISA, Taiji, and TianQin experiments.
    We have also checked that the DM candidate $\chi$ of these models has a total annihilation cross section 
    $\langle\sigma v\rangle_{\rm tot}\simeq 2.2\times 10^{-26}$~$\rm cm^3/s$ (see Table~\ref{tab:results}), which is much lower than the upper 
    limits on the DM annihilation to SM particles from Fermi-LAT~\cite{Fermi2015PRL} and MAGIC~\cite{MAGIC2016JCAP} gamma-ray observations of 
    dwarf satellite galaxies. }
    \label{fig:gw}
\end{figure}
\begin{table} 
    \centering
    \caption{Summary of scenarios used in the right plot Fig.~\ref{fig:gw}, assuming $\lambda_{\chi}=0.5$ and $\kappa_{\chi}=0$. }
    \begin{tabular}{cccccccccccccc}
      \hline
      \hline
        Scenario & $m_S$/GeV & $m_{\chi}$/GeV & $\theta$ & $\lambda_a$ & $c_{\chi}$/GeV & $\lambda_s$ & $\kappa_s$ & $c_s$/GeV & $\lambda_h$ \\
      \hline
      A & 182.0 & 1524.0 & 0.342 & 0.467 & $-403.0$ & 1.3 & 0.21 & $-367.0$ & 0.145 \\
      B & 109.0 & 984.0 & 0.279 & 0.309 & 162.0 & 0.71 & 0.451 & $-135.0$ & 0.127 \\
      C & 137.0 & 1887.1 & 0.228 & 0.581 & $-478.0$ & 0.85 & 0.403 & $-200.0$ & 0.130 \\
      \hline
      \hline
    \end{tabular}
    \label{tab:models}
\end{table} 
\begin{table} 
    \centering
    \caption{Summary of results for the three scenarios given in Table~\ref{tab:models}.}
    \begin{tabular}{cccccccccccccc}
      \hline
      \hline
      Scenario & $\sigma_{\rm SI}$/${\rm cm^2}$ & $\langle\sigma v\rangle_{\rm tot}$/${\rm cm^3s^{-1}}$ & $v_c/T_c$ & $T_c$/GeV & $w_c$/GeV & $T_n$/GeV & $\beta/H_{*}$ & $\alpha$ \\
      \hline
      A & $1.27\times 10^{-46}$ & $2.17\times 10^{-26}$ & 1.32 & 173.0 & 194.7 & 53.57 & 1343.0 & 0.139 \\
      B & $1.23\times 10^{-47}$ & $2.16\times 10^{-26}$ & 1.10 & 196.3 & 186.2 & 55.96 & 1680.8 & 0.112 \\
      C & $5.70\times 10^{-48}$ & $2.17\times 10^{-26}$ & 1.13 & 193.4 & 192.6 & 38.37 & 512.8 & 0.523 \\
      \hline
      \hline
    \end{tabular}
    \label{tab:results}
\end{table} 

We present our main results in Fig.~\ref{fig:gw}.  The colored regions of the left plot denote the detectable parameter spaces for LISA (orange), 
B-DECIGO (light-green), DECIGO (light-blue), and BBO (pink).  
As illustrated above, we follow the scenario provided in Ref.~\cite{Chiang2020JHEP} to search for the first-order EWPT 
for the one-loop potential~\eqref{eq:potentot} and take into account the constraints from vacuum stability, perturbativity, and perturbative 
unitarity as well as the Higgs signal strength observations, as done in Refs.~\cite{Chiang2020JHEP,Chiang2021JCAP}.
We then check the bubble nucleation criterion and calculate the GW parameters with the help of 
{\tt CosmoTransitions 2.0.2}~\cite{CosmoTransitions2012}.  Furthermore, the DM phenomenology including the DM relic density as well as direct 
detection cross sections has also been investigated using the {\tt MicrOMEGAs 5.0.4} package~\cite{MicrOMEGAs2018CPC} in which the model is 
implemented with the {\tt FeynRules 2.3} package~\cite{FeynRules2014CPC}.  All the scattering points depicted in the left plot further have the correct DM relic density $h^2\Omega_{\rm DM}\simeq 0.12$ as obtained by the Planck 
observations~\cite{Planck2016} in the $\Lambda$CDM framework. More interestingly, some of the above DM benchmarks with the correct DM relic 
density can even evade the current stringent constraints from DM direct and indirect searches.
We select three scenarios with their parameters given in Table~\ref{tab:models}, while the numerical predictions of some important DM and GW 
quantities for each model are presented in Table~\ref{tab:results}.  The GW signals from the first-order EWPT in these selected scenarios are 
depicted in the right plot of Fig.~\ref{fig:gw}.  We observe that these signals have both the appropriate peak frequencies and the  sufficient 
amplitudes, so they are suitable targets for LISA~\cite{LISA2017,Robson2019CQG}, Taiji~\cite{Hu2017NSR,Ruan2020NA}, 
TianQin~\cite{Luo2016CQG,Hu2018CAG}, BBO~\cite{Crowder2005PRD}, and U-DECIGO~\cite{Kudoh2006PRD} experiments. Therefore, 
as demonstrated with this simple complex-singlet-scalar extended model, 
the future space-based GW observations provide us an important new tool to probe the DM physics.

{\it Summary -- }
Although the WIMP with mass around 1~TeV is one of the most popular DM scenarios, in recent years it suffers from grave challenges from DM 
direct detections, cosmic ray observations, and collider searches. In fact, previous works have shown that the singlet extensions of Higgs 
portal WIMP models with a $Z_{2}$ or $U(1)$ symmetry have already been well constrained by the XENON1T experiment as well as the Fermi-LAT 
gamma-ray measurements. We have found in Ref.~\cite{Chiang2021JCAP} that by extending to the $CP$ symmetry, it is possible for the singlet extension model to trigger a strong first-order EWPT and to provide a correct DM relic abundance while evading various constraints from DM experiments.  In this work, we focus on studying the stochastic gravitational-wave signals emitted from a strong first-order EWPT in such a model. We show that there are ample parameter samples that can give rise to a correct DM relic abundance while generating a significant stochastic GW background, which can be probed by near-future space-based GW experiments, including BBO, U-DECIGO, LISA, Taiji, and TianQin.  
Among these model samples, we identify three benchmarks with the DM mass $\sim 1$~TeV and the total annihilation cross section $\simeq 2\times 10^{-26}$~${\rm cm^3/s}$.  Two of them are testable by the future XENONnT underground DM experiment, while the GW signals from all three benchmarks are sufficiently loud for upcoming space-based GW observations.  By this simple model example, we hope to demonstrate that the GW measurement provides us a new way to detect and probe the nature of DM, complementary to the traditional detection methods.

{\it Acknowledgments -- }
BQL is supported in part by the Huzhou University under startup Grant No. RK21094 and National Natural Science Foundation of China (NSFC) under Grant No. 12147219. CWC is supported in part by the Ministry of Science and Technology of Taiwan under Grant No.~MOST-108-2112-M-002-005-MY3.
DH is supported in part by the National Key Research and Development Program of China under Grant No. 2021YFC2203003, National Natural Science Foundation of China (NSFC) under Grant No. 12005254 and the Key Research Program of Chinese Academy of Sciences under grant No. XDPB15.

\end{document}